\documentstyle[11pt,fleqn]{article}  
\input{psfig.tex}                
\title{Dispersion Relations for Waves \\ propagating in Composite Fermion Gases} 
\author{A. Kwang-Hua Chu} 
\date{Department of Physics, Xinjiang University,
Urumqi 830046, PR China}
\begin{document}      
\maketitle  
\begin{abstract}           
The discrete Uehling-Uhlenbeck equations  are solved to study the
propagation of plane (sound) waves in a system of composite
fermionic particles with hard-sphere interactions and the filling
factor ($\nu$) being 1/2. The Uehling-Uhlenbeck collision sum, as
it is highly nonlinear, is linearized firstly and then decomposed
by using the plane wave assumption. We compare the dispersion
relations thus obtained by the relevant Pauli-blocking parameter
$B$ which describes the different-statistics particles for the
quantum analog of the discrete Boltzmann system when $B$ is
positive (Bose gases), zero (Boltzmann gases), and negative (Fermi
Gases). We found, as the effective magnetic field being zero
($\nu$=1/2 using the composite fermion formulation), the electric
and fluctuating (induced) magnetic fields effect will induce anomalous dispersion relations.
\newline

\noindent  %
PACS : \hspace*{2mm} 71.10.Pm, 73.43.-f  \newline
\end{abstract}
%
\doublerulesep=6mm    
\baselineskip=6mm  
\bibliographystyle{plain}
\section{Introduction}
The study of the electronic properties of quasi-two-dimensional
(2D) systems has resulted in a number of remarkable discoveries in
the past two decades [1-6]. Among the most interesting of these
are the integral and fractional quantum Hall effects [1] (the
integral quantum Hall effect, which is manifested by the
development of spectacularly flat plateaus in the Hall conductance
centered at integral values of $n$, was discovered in 1980 by
Klaus von Klitzing). In both of these effects, incompressible
states of a 2D electron liquid are found at particular values of
the electron density for a given value of the magnetic field
applied normal to the 2D layer.
\newline In the presence of a strong magnetic field B transverse
to a two-dimensional system of electrons, the tiny cyclotron
orbits of an electron are quantized to produce discrete kinetic
energy levels, called {\it Landau levels}.  The degeneracy of each
Landau level-that is to say, its maximum population per unit
area-is ${\cal B}$/$\phi_0$, where $\phi_0 = h/e$ is the
elementary quantum of magnetic flux. This degeneracy implies that
the number of occupied Landau levels, called the filling factor,
is $\nu = \rho \phi_0$ /${\cal B}$, where $\rho$ is the
two-dimensional electron density.
\newline
The fractional quantum Hall effect (FQHE) is more difficult to
understand and more interesting in terms of new basic physics. The
energy gap that gives rise to the Laughlin [2] incompressible
fluid state is completely the result of the interaction between
the electrons. The elementary excitations are fractionally charged
Laughlin quasiparticles, which satisfy fractional statistics [2].
The standard techniques of many-body perturbation theory are
incapable of treating FQH systems because of the complete
degeneracy of the single-particle levels in the absence of the
interactions. Laughlin [2] was able to determine the form of the
ground-state wavefunction and of the elementary excitations on the
basis of physical insight into the nature of the many-body
correlations. Striking confirmation of Laughlin＊s picture was
obtained by exact diagonalization of the interaction Hamiltonian
within the subspace of the lowest Landau level of small systems
[2]. Jain, Lopez and Fradkin,and Halperin {\it et al.} [3-4] have
extended Laughlin＊s approach and developed a composite-fermion
(CF) description of the 2D electron gas in a strong magnetic
field.  The composite-fermion (CF) picture offers a simple
intuitive way of understanding many of the surprising properties
of a strongly interacting two-dimensional electron fluid in a
large magnetic field. 
\newline
The quickest way to introduce the composite fermion is through the
following series of steps, which Jain called the {\it Bohr theory}
of composite fermions because it obtains some of the essential
results with the help of an oversimpified but useful picture [7].
The outcome is that strongly interacting electrons in a strong
magnetic field ${\cal B}$ transform into weakly interacting
composite fermions in a weaker effective magnetic field ${\cal
B}_{\mbox{eff}}$, given by ${\cal B}_{\mbox{eff}}$ =${\cal
B}-2p\,\phi_0 \rho$, where $2p$ is an even integer. Equivalently,
one can say that electrons at filling factor n convert into
composite fermions with filling factor $\nu^* = \rho \phi_0
/|{\cal B}_{\mbox{eff}}|$, given by
\begin{displaymath}
  \nu=\frac{\nu^*}{2p \,\nu^* \pm 1}.
\end{displaymath}
The minus sign corresponds to situations when ${\cal
B}_{\mbox{eff}}$ points antiparallel to ${\cal B}$. Start by
considering interacting electrons in the transverse magnetic field
${\cal B}$. Now attach to each electron an infinitely thin,
massless magnetic solenoid carrying 2p flux quanta pointing
antiparallel to ${\cal B}$, turning it into a composite fermion.
Such a conversion preserves the minus sign associated with an
exchange of two fermions, because the bound state of an electron
and an even number of flux quanta is itself a fermion. Hence the
name. It also leaves the Aharonov-Bohm phase factors associated
with all closed paths unchanged, because the additional phase
factor due to a flux $\phi = 2p \phi_0$ is exp$[2i\pi\phi/\phi_0]
= 1$. In other words, the attached flux is unobservable, and the
new problem, formulated in terms of composite fermions, is
identical to the one with which we began.\newline
The crucial point is that the many-particle ground state of
electrons at $\nu < 1$ was highly degenerate in the absence of
interaction, with all lowest Landau level configurations having
the same energy. But now, the degeneracy of the composite-fermion
ground state at the corresponding $\nu^* > 1$ is drastically
smaller, even when the interaction between composite fermions is
switched off. For integral values of $\nu^*$, in fact, one gets a
non-degenerate ground state. The reduced degeneracy suggests that
one might start by treating the composite fermions as independent.
In that approximation, the composite fermions fill a Fermi sea of
their own whenever ${\cal B}_{\mbox{eff}}$ vanishes ($1/\nu =
2p$), and form composite-fermion Landau levels when it does not.
\newline
Composite fermions (CF), consisting of an electron with two flux
quanta attached, provide a different approach to the fractional
quantum Hall effect (FQHE) [3]. At filling factor $\nu$ = 1/2 the
attached flux quanta are `compensated' by externally applied
magnetic flux such that the CF move in a vanishing effective
magnetic field. Away from  $\nu$= 1/2 the effective magnetic field
increases, the CF move on circles with radius R$_{C;CF}$ and the
Landau quantization of the circular motion of the new particles is
the origin of the FQHE. The radius R$_{C;CF}$ is given by $\hbar
\sqrt{4\pi n_s}/e\, {\cal B}_{\mbox{eff}}$ with the electron
density $n_s$, the effective magnetic field ${\cal B}_{\mbox{eff}}
={\cal B} - {\cal B}_{1/2}$, and ${\cal B}_{1/2}$ the magnetic
field at $\nu$ = 1/2 . Experimental evidence for the existence of
CF mainly stems from commensurability experiments where the Fermi
wave vector $k_{F;CF}$ of the novel quasi-particles is probed by a
periodic external perturbation.
\newline
An important application of the concept concerns the metallic
state at $\nu = 1/2$ , where no fractional quantum Hall state is
seen. If composite fermions exist at that filling factor, they
would experience no effective magnetic field (${\cal
B}_{\mbox{eff}} = 0$). Thus a mean-field picture suggests a Fermi
sea of composite fermions.
\newline
Although this CF description has offered a simple picture for the
interpretation of many experimental results. However, the
underlying reason for the validity of many of the approximations
used with the CF approach is not completely understood [7-8].
\subsection{Previous Semi-Classical Approaches}
A semi-classical theory based on the Boltzmann transport equation
for a two-dimensional electron gas modulated along one direction
with weak electrostatic or magnetic modulations have been proposed
[9-11]. Ustinov and Kravtsov studied the giant magnetoresistance
effect in magnetic superlattices for the current perpendicular to
and in the layer planes within a unified semiclassical approach
that is based on the Boltzmann equation with exact boundary
conditions for the spin-dependent distribution functions of
electrons. Interface processes responsible for the
magnetoresistance were found to be different in these geometries,
and that can result in an essential difference in general
behaviour between the in-plane magnetoresistance and the
perpendicular-plane one. A correlation between the giant
magnetoresistance and the multilayer magnetization is also
discussed therein [9]. \newline Boltzmann's equation provides an
adequate starting point of transport calculations for
two-dimensional electron systems in the presence of periodic
electric and magnetic modulation fields, both in the regime of the
low-field positive magnetoresistance and of the Weiss oscillations
at intermediate values of the applied magnetic field. For example,
Zwerschke and Gerhardts solved Boltzmann's equation by the method
of characteristics, which allows to exploit explicitly information
about the structure of the phase space. That structure becomes
very complicated if the amplitudes of the modulation fields become
so large and the average magnetic field becomes so small that, in
addition to the drifting cyclotron orbits, channeled orbits exist
and drifting cyclotron orbits extend over many periods of the
modulation [10]. \newline
In Refs. [5,10], they considered the 2DEG in the $x$-$y$ plane as
a degenerate Fermi gas, with Fermi energy $E_F=m^*\,v_F^2/2$, of
(non-interacting) particles with effective mass $m^*$ and charge
$-e$ obeying classical dynamics, i.e. Newton's equation $m^*
\dot{\bf v}$= $-e[{\bf F}+({\bf v}\times {\cal B})]$. In
equilibrium, the electric field is given by ${\bf F}({\bf
r})=\nabla V({\bf r})/e$, where $V({\bf r})$ is the modulating
electrostatic potential. In thermal equilibrium, all states with
energy below $E_F$ are occupied, and for the linear response to an
external homogeneous electric field ${\bf E}_0$ only the electrons
with energy $E({\bf r},{\bf v})=(m^* \,{\bf v}^2)/2+ V({\bf
r})=E_F$ contribute to the current. The distribution function
$f({\bf r},{\bf v},t)$ obeys the Boltzmann equation
\begin{displaymath}
 \frac{\partial f}{\partial t}+{\cal D} f-{\cal C} [f;{\bf r},{\bf v}]=
 {\bf v}\cdot {\bf E}_0,
\end{displaymath}
where the drift term ${\cal D}$ describes the change due to the
natural motion of the electrons in the modulation field (in
absence of ${\bf E}_0$), and ${\cal C}$ is the collision operator.
We might use polar coordinates in the velocity space, ${\bf v}= v
{\bf u}$ with $v({\bf r})=v_F[1-V({\bf r})/E_F]^{1/2}$  and ${\bf
u}(\Theta$)=($\cos \Theta, \sin \Theta$). Sometimes [2], the drift
term  reads ${\cal D}={\bf v} \cdot \nabla+ [\omega_c +\omega_{el}
({\bf r},\Theta)]\partial/\partial \Theta$,  with cyclotron
frequency $\omega_C=e {\cal B}_{\mbox{eff}}/m^*$ and
$\omega_{el}({\bf r},\Theta)=(\nabla V) {\bf t}$ with ${\bf
t}(\Theta)=(\sin \Theta, \cos \Theta)$.
\newline
Recently Jobst investigated the magnetoresistance of a weakly
density modulated high mobility two-dimensional electron system
around filling factor $\nu$ = 1/2 [12]. The experimental
$\rho_{xx}$-traces around $\nu$ = 1/2 were well described by novel
model calculations, based on a semiclassical solution of the
Boltzmann equation, taking into account anisotropic scattering. We
also noticed that, the effects of a tunable periodic density
modulation imposed upon a 2D electron system have been probed
using surface acoustic waves by Willett {\it et al.} [13]. A
substantial effect was induced at filling factor 1/2 in which the
Fermi surface properties of the CF are anisotropically replaced by
features similar to those seen in quantum Hall states. The
response measured using different SAW wavelengths and similarities
in the temperature dependence between the modulation induced
features at 1/2 and quantum Hall states were described therein
[13].
\subsection{Present Objectives}
Motivated by the interesting issues about $\nu$=1/2, we like to
study their characteristics relevant to the sound propagation in
CF gases here using our verified quantum (discrete) kinetic
approaches [14-15].  
In the discrete kinetic model approach [16], the main idea is to
consider that the particle velocities belong to a given finite set
of velocity vectors, e.g.,  ${\bf u}_1, {\bf u}_2, \cdots, {\bf
u}_p$, $p$ is a finite positive integer. Only the velocity space
is discretized, the space and time variables are continuous
[15-16] (please see the detailed references therein). By using the
discrete velocity model approach, the velocity of propagation of
plane waves can be classically determined by looking for the
properties of the solution of the conservation equation referred
to the equilibrium state. \newline As a continuous attempt of
ultrasonic propagation (in dilute  Boltzmann gases [17]),
considering the quantum analog of the discrete velocity model and
the Uehling-Uhlenbeck collision term which could describe the
collision of a gas of dilute hard-sphere Fermi-, Boltzmann- or
Bose-particles by tuning a parameter $\theta$ [14,18] (via a {\it
blocking factor} of the form $1+\theta f$ with $f$ being a
normalized distribution function giving the number of particles
per cell, say, a unit cell, in phase space), in this paper, we
plan to investigate the dispersion relations of plane ultrasonic
waves propagating in composite-fermion gases by the quantum
discrete kinetic model which has been verified before. The CF-CF
interactions [6,8] will not be considered in present works  since
our present approach works quite well only for the dilute
(weakly-interacting) regime
 [15-20]. This presentation will give more
clues to the studies of the quantum wave dynamics in composite
fermion gases [13].
\section{Mathematical Formulations}
The gas is presumed to be composed of identical hard-sphere
particles of the same mass. The discrete number density (of
particles) is denoted by $N_i ({\bf x},t)$ associated with the
velocity ${\bf u}_i$ at point ${\bf x}$ and time $t$. Following
the CF model, around $\nu$ = 1/2 or any even-denominator $\nu =
1/2p$, $2p$ fictitious magnetic flux quanta ($\phi_0 = h/e$) are
attached to each electron in the direction opposite to the
external magnetic field ${\cal B}$. The so formed composite particles
follow Fermi statistics and are named composite fermions. The flux
attachment transforms the strongly interacting two-dimensional
electron system (2DES) of density $\rho$ in a high a magnetic
field into an equivalent weakly interacting CF system, which
experiences a smaller effective magnetic field, ${\cal
B}_{\mbox{eff}} ={\cal B}- 2\rho p \phi_0$. In particular, at
exact even-denominator fillings, $\nu = 1/2p$, ${\cal B}=2p\,\rho
\,h/e = 2\rho p\,\phi_0$ and ${\cal B}_{\mbox{eff}}$ vanishes.
Under these conditions, the CFs reside in a magnetic field-free
region and, like ordinary 2D electrons at ${\cal B}$ = 0, they
form a Fermi sea. 
\newline If only nonlinear binary collisions and the effective
magnetic field ${\bf B}_{\mbox{eff}}$ being zero (for $\nu=1/2$ in
the CF sense) are considered,   we have for the evolution of
$N_i$,
\begin{equation}
 \frac{\partial N_i}{\partial t}+ {\bf u}_i \cdot \nabla N_i
 -\frac{e ({\bf E}+{\bf u}_i \times {\bf B}_{\mbox{eff}})}{m^*} \cdot \nabla_{ {\bf u}_i} N_i=
 {\cal C}_i \equiv \sum^p_{j=1} \sum_{(k,l)} (A^{ij}_{kl} N_k N_l - A^{kl}_{ij}
 N_i N_j),  \hspace*{3mm} i=1,\cdots, p,
\end{equation}
where ${\bf E}$ is the electric field, $m^*$ is the effective mass
of the particle, $(k,l)$ are admissible sets of collisions
[14-18]. We may also define the right-hand-side of above equation
as
\begin{equation}
 {\cal C}_i (N) =\frac{1}{2}\sum_{j,k,l} (A^{ij}_{kl} N_k N_l - A_{ij}^{kl}
 N_i N_j),
\end{equation}
with $i \in$ $\Lambda$ =$\{1,\cdots,p\}$, and the summation is
taken over all $j,k,l \in \Lambda$, where $A_{kl}^{ij}$ are
nonnegative constants satisfying [14-18] (i)
  $A_{kl}^{ji}=A_{kl}^{ij}=A_{lk}^{ij}$ : {\it indistinguishability of the
  particles in collision},
(ii) $A_{kl}^{ij} (u_i +u_j -u_k -u_l )=0$ :  {\it conservation of
momentum in the collision},
(iii) $A_{kl}^{ij}=A_{ij}^{kl}$ : {\it microreversibility
condition}.
The conditions defined for discrete velocities above are valid for
elastic binary collisions such that momentum and energy are
preserved.
The collision operator is now simply obtained by joining
$A_{ij}^{kl}$ to the corresponding transition probability
densities $a_{ij}^{kl}$ through $ A_{ij}^{kl}$ =$ S|{\bf u}_i-{\bf
u}_{j}|$ $a_{ij}^{kl}$, where,
\begin{displaymath}
 a_{ij}^{kl} \ge 0 , \hspace*{12mm} \sum^p_{k,l=1}  a_{ij}^{kl}=1 ,
 \hspace*{3mm} \forall \,i,j=1,\cdots,p ;
\end{displaymath}
with $S$ being the effective collisional cross-section [14-18]. If
all $n$ ($p=2 n$) outputs are assumed to be equally probable, then
$a_{ij}^{kl}$=$1/n$ for all $k$ and $l$, otherwise $a_{ij}^{kl}$=
0.
Collisions which satisfy the conservation and reversibility
conditions which have been stated above are defined an {\it
admissible collision} [14-18]. 

\noindent With the introducing of the Uehling-Uhlenbeck collision
term [18] in Eq. (1) or Eq. (2),
\begin{equation}
 {\cal C}_i =\sum_{j,k,l} A^{ij}_{kl} \,[ N_k N_l (1+\theta N_i)(1+\theta N_j)-
 N_i N_j (1+\theta N_k)(1+\theta N_l)],
\end{equation}
for $\theta <0$ we obtain a gas of Fermi-particles; for $\theta
> 0$ we obtain a gas of Bose-particles, and for $\theta =0$ we obtain
Eq. (1).  \newline From Eq. (3), the model of discrete quantum
Boltzmann equation for dilute hard-sphere gases proposed in [18]
is then a system of $2n(=p)$ semilinear partial differential
equations of the hyperbolic type :
\begin{displaymath}
 \frac{\partial}{\partial t}N_i +{\bf v}_i \cdot\frac{\partial}{\partial
 {\bf x}} N_i -\frac{e ({\bf E}+{\bf v}_i \times {\bf B}_{\mbox{eff}})}{m^*} \cdot \nabla_{{\bf v}_i} N_i
 =\frac{c S}{n} \sum_{j=1}^{2n} N_j N_{j+n}(1+\theta N_{j+1})(1+\theta N_{j+n+1})
\end{displaymath}
\begin{equation}
 -
 2 c S N_i  N_{i+n} (1+\theta N_{i+1})(1+\theta
 N_{i+n+1}), 
\end{equation}
where $N_i=N_{i+2n}$ are unknown functions, and ${\bf v}_i$ =$ c
(\cos[(i-1) \pi/n], \sin[(i-1)\pi/n])$, $i=1,\cdots, 2 n$; $c$ is
a reference velocity modulus [14-18]. 
The admissible collisions as $n=2$ are $({\bf v}_1,{\bf v}_{3})
\longleftrightarrow ({\bf v}_2,{\bf v}_{4})$. \newline We notice
that the right-hand-side of the Eq. (4) is highly nonlinear and
complicated for a direct analysis. As passage of the sound wave
causes a small departure from an equilibrium resulting in energy
loss owing to internal friction and heat conduction, we linearize
above equations around a uniform equilibrium state ($N_0$) by
setting $N_i (t,x)$ =$N_0$ $(1+P_i (t,x))$, where $P_i$ is a small
perturbation. The equilibrium here is presumed to be the same as
in Refs. [14-15,18](in the absence of applied fields, the
electrons will be at equilibrium and the distribution function
will be the equilibrium distribution function $N_0
(\epsilon-\mu_0)= [1+\exp(\epsilon-\mu_0)/k_B\,T)]^{-1}$, where
$\mu_0$ is the chemical potential, $k_B$ is the Boltzmann
constant, the corresponding Fermi surface is defined by the
equations $\epsilon({\bf k})=\mu_0$ in the quasi-momentum space,
${\bf k}$ is the wave vector). After some similar manipulations as
mentioned in Refs. [15,17], with $B=\theta N_0$ [14-15], which
gives or defines the (proportional) contribution from dilute Bose
gases (if $\theta
> 0$, e.g., $\theta=1$), or dilute Fermi gases (if $\theta < 0$,
e.g., $\theta=-1$), we then have
\begin{equation}
 [\frac{\partial^2 }{\partial t^2} +c^2 \cos^2\frac{(m-1)\pi}{n}
 \frac{\partial^2 }{\partial x^2} +4 c S N_0 (1+B) \frac{\partial
 }{\partial t}] D_m- \frac{4 c S N_0 (1+B)}{n} \sum_{k=1}^{n} \frac{\partial
 }{\partial t} D_k ={\mbox{RHS}} ,
\end{equation}
where $D_m =(P_m +P_{m+n})/2$, $m=1,\cdots,n$, since $D_1 =D_m$
for $1=m$ (mod $2 n)$. Here, RHS denotes the contribution from the
electric field and the fluctuating induced magnetic field. This term could be worked out by following the
previous approaches [9,21] (cf. the second term in the left-hand
side of the equation (4) in [9]).
\newline
We are ready to look for the solutions in the form of plane wave
$D_m$= $d_m$ exp $i (k x- \omega t)$, $(m=1,\cdots,n)$, with
$\omega$=$\omega(k)$. This is related to the dispersion relations
of (forced) plane waves propagating in dilute (monatomic)
hard-sphere Bose ($B>0$) or Fermi ($B<0$) gases. So we have
\begin{equation}
 (1+i h (1+B)-2 \lambda^2 cos^2 \frac{(m-1)\pi}{n}) d_m -\frac{i h (1+B)}{n}
 \sum_{k=1}^n d_k = {\mbox{RHS}} , \hspace*{10mm} m=1,\cdots,n,
\end{equation}
with
\begin{displaymath}
\lambda=k c/(\sqrt{2}\omega),  \hspace*{12mm} h=4 c S N_0/\omega,
\end{displaymath}
where $\lambda$ is complex and $h \,(\propto 1/\mbox{Kn})$ is the
rarefaction parameter of the Bose- or Fermi-particle gas (Kn is
the Knudsen number which is defined as the ratio of the mean free
path of Bose or Fermi gases to the wave length of the plane
(sound) wave).
\subsection{Weak External Fields}
We firstly consider the case of rather weak electric field together with rather weak
fluctuating (induced) magnetic field. It
means  RHS $\approx 0$ considering other domainted terms in the
equation (6). Let $d_m$ = ${\cal{C}}/(1+i h (1+B)-2 \lambda^2
\cos^2 [(m-1)\pi/n])$, where ${\cal{C}}$ is an arbitrary, unknown
constant, since we here only have interest in the eigenvalues of
above relation. The eigenvalue problems for
different $2\times n$-velocity model reduces to 
\begin{equation}
  1-\frac{i h (1+B)}{n} \sum^n_{m=1} \frac{1}{1+i h (1+B)-2 \lambda^2
  \cos^2\,\frac{(m-1)\pi}{n}} ={\mbox{RHS}} \sim 0.
\end{equation}
We solve only $n=2$ case, i.e., 4-velocity case since for $n>2$
there might be spurious invariants [14-15]. 
For 2$\times$2-velocity model, we obtain
 $$1-[i h(1+B)/2] \sum^2_{m=1} \{1/[1+i h(1+B)-2 \lambda^2  \cos^2\,(m-1)\pi/2 ]\}
 =0. $$ 
\section{Results and Discussions}
With the filling factor $\nu$=1/2, we are now ready to obtain the
dispersion relations for sound propagating in composite fermion
gases (with ${\cal B}_{\mbox{eff}}$=0) which might be useful to
those subsequent studies reported in [13] by using surface
acoustic waves. By using the standard symbolic or numerical
software, e.g. Mathematica or Matlab, we can obtain the complex
roots ($\lambda=\lambda_r +$ i $\lambda_i$) from the polynomial
equation above. The roots are the values for the
nondimensionalized dispersion (positive real part; a relative
measure of the sound or phase speed) and the attenuation or
absorption (positive imaginary part), respectively.
\newline Curves in Fig. 1 or 2 follow the conventional dispersion
relations of ultrasound propagation in dilute hard-sphere
(Boltzmann; $B=0$) gases [17,19-20]. Here, $s$-scattering means
the conventional $s$-wave scattering. Our results show that as
$|B|$ ($B$: the Pauli-blocking parameter) increases, the
dispersion ($\lambda_r$) will reach the continuum or
hydrodynamical limit ($h \rightarrow \infty$) earlier. The phase
speed of the plane (sound) wave in Bose gases (even for small but
fixed $h$) increases more rapid than that of Fermi gases (w.r.t.
to the standard conditions : $h \rightarrow \infty$) as the
relevant parameter B increases. For all the rarefaction measure
($h$), plane waves propagate faster in Bose-particle gases than
Boltzmann-particle and Fermi-particle gases. Meanwhile, the
maximum absorption (or attenuation) for all the rarefaction
parameters $h$ keeps the same for all $B$ as observed in Fig. 2.
There are only shifts of the maximum absorption state (defined as
$h_{max}$) w.r.t. the rarefaction parameter $h$ when $B$
increases.  It seems for the same mean free path or mean collision
frequency of the dilute hard-sphere gases (i.e. the same $h$ as
$h$ is small enough but $h < h_{max}$) there will be more
absorption in Bose particles than those of Boltzmann and Fermi
particles when the plane (sound) wave propagates.
\newline On the contrary, for the same $h$ (as $h$ is large enough
but $h> h_{max}$, there will be less absorption in Bose particles
than those of Boltzmann particles when the plane wave propagates.
When $B$ (i.e., $\theta$) is less than zero or for the
Fermi-particle gases, the resulting situations just mentioned
above reverse. For instance, as the rarefaction parameter is
around 10, which is near the hydrodynamical or continuum limit, we
can observe that the ultrasound absorption becomes the largest
when the plane (sound) wave propagates in hard-sphere Fermi gases.
That in Bose gases becomes the smallest. As also illustrated in
Fig. 1 for cases of dilute Fermi gases ($B<0$), the rather small
dispersion value (relative measure of different phase speeds
between the present rarefied state : $h$ and the hydrodynamical
state : $h \rightarrow \infty$) when $B$ approaches to $-1$
perhaps means there is the Fermi pressure which causes a Fermi gas
to resist compression.
\newline From a modern point of view, dissipations of the (forced)
plane (sound) wave arise fundamentally because of a necessary
coupling between density and energy fluctuations induced by
disturbances. Within one mean free path or so of an oscillating
boundary, a free-particle flow solution can probably be computed.
The damping will quite likely turn out to be linear because the
damping mechanism is the shift in phase of particles which hit the
wall at different times.  As the wavelength is made significantly
shorter, so that the effects of viscosity and the heat conduction
are no longer small, the validity of hydrodynamic approach itself
becomes questionable. If there is no rarefaction effect ($h=0$),
we have only real roots for all the models. Once $h \not=0$, the
imaginary part appears and the spectra diagram for each gas looks
entirely different. In short, the dispersion ($k_r
c/(\sqrt{2}\omega)$) reaches a continuum-value of $1$ (or
saturates) once $h$ increases to infinity. We noticed that the
increasing trend for the expression of our dispersion
($\lambda_r$; dimensionless) when waves propagating in Bose gases
is similar to that (of dimensional sound speed) reported in Ref.
[22-23]. The absorption or attenuation ($k_i c/(\sqrt{2}\omega)$)
for our model, instead, firstly increases up to $h\sim 1$,
depending upon the $B$ values, then starts to decrease as $h$
increases furthermore. \newline Although curves of the dispersion
relation for hard-sphere Bose gases resemble qualitatively those
reported in Refs. [22-23]. But, because of many unknown baselines
(for example, in Fig. 1 of Ref. [22], their horizontal axis is
represented by the condensate peak density which may be linked to
our rarefaction parameter, however, at present, the detailed link
is not available), we cannot directly compare ours with their
data. The results presented here also show the intrinsic
thermodynamic properties of the equilibrium states corresponding
to the final equilibrium state after the collision of dilute
hard-sphere Bose ($B>0$), Boltzmann ($B=0$), and Fermi ($B<0$)
gases. \newline At low temperatures, the Pauli exclusion principle
forces Fermi-gas particles to be farther apart than the range of
the collisional interaction, and they therefore cannot collide and
rethermalize. That is to say, identical fermions are unable to
undergo the collisions necessary to rethermalize the gas during
evaporation  because of the need to maximize Pauli blocking efects
[15]. The much more spreading characteristics of dispersion
relations for dilute Fermi gases ($B <0$) obtained and illustrated
in Figs. 1 and 2 seems to confirm above theoretical reasoning. The
deviations in curves of dispersion and absorption shown in Figs. 1
and 2 also highlight their dissimilar quantum statistical nature.
\newline
Considering the case of nonzero electric and fluctuating (induced) magnetic fields, i.e.,
RHS $\not
=0$, we can obtain the detailed mathematical expression for RHS by
following the verified approaches [9,21] with
\begin{equation}
 \mbox{RHS}\equiv i \frac{e (|{\bf E}+{\bf v}_i \times {\cal B}_{\mbox eff}|)}{m^*} \delta(\epsilon-\mu_0) c
 \cos(\frac{m-1}{n}\pi)[c  \cos(\frac{m-1}{n}\pi) k+\omega],
 \end{equation}
where $\delta$ is the delta function. To obtain similar dispersion
relations together with the equation (6) or (7) with nonzero RHS,
we must impose the other condition from the equation (8) with RHS
being zero for arbitrary ${\cal C}$. Under this situation, we have
anomalous results : $|\lambda_r|=1/\sqrt{2}$ ($\lambda_r$ is
negative!) and $\lambda_i =0$ for all the rarefaction measure
($h$s) and the Pauli-blocking parameter ($B$s). This strange
behavior for $\nu$=1/2 (${\cal B}_{\mbox{eff}}=0$, the electric
field (${\bf E}$) effect is being considered or both external fields are present) within the composite
fermion formulation, however, is similar to that reported in [15]
for the specific case of sound propagating in normal fermionic
gases (the Pauli-blocking parameter $B=-1$) or sound propagating
in dilute gases (for all $B$s but with a free orientation
parameter being $\pi/4$). There is no attenuation for above
mentioned cases. This last observation might be relevant to the
found {\it enhanced} conductivity (for 2D electron gases)
corresponding to the even-denominator factor $\nu$=1/2 (composite
fermions) using surface acoustic waves (of wavelength smaller than
1 $\mu$m) [13] (geometric resonance of the composite fermions'
cyclotron orbit and the ultrasound wavelength was also observed at
smaller wavelength therein). \newline To conclude in brief, by
using the quantum discrete kinetic approach, for the case of
nonzero electric field, we obtain strange dispersion relations for
waves propagating in CF gases with $\nu=1/2$ :
$|\lambda_r|=1/\sqrt{2}$ ($\lambda_r$ is negative) and $\lambda_i
=0$ for all the rarefaction measure ($h$s) and the Pauli-blocking
parameter ($B$s). We shall investigate other interesting issues
(e.g., compressibility of CF [24-25]) in the future.
{\small Acknowledgements.  The author is
partially supported by
the Starting Funds for the 2005-XJU-Scholars.}

\newpage
\vspace{30mm} \oddsidemargin=0mm

\pagestyle{myheadings}

\topmargin=-10mm

\textwidth=17cm \textheight=25cm
\psfig{file=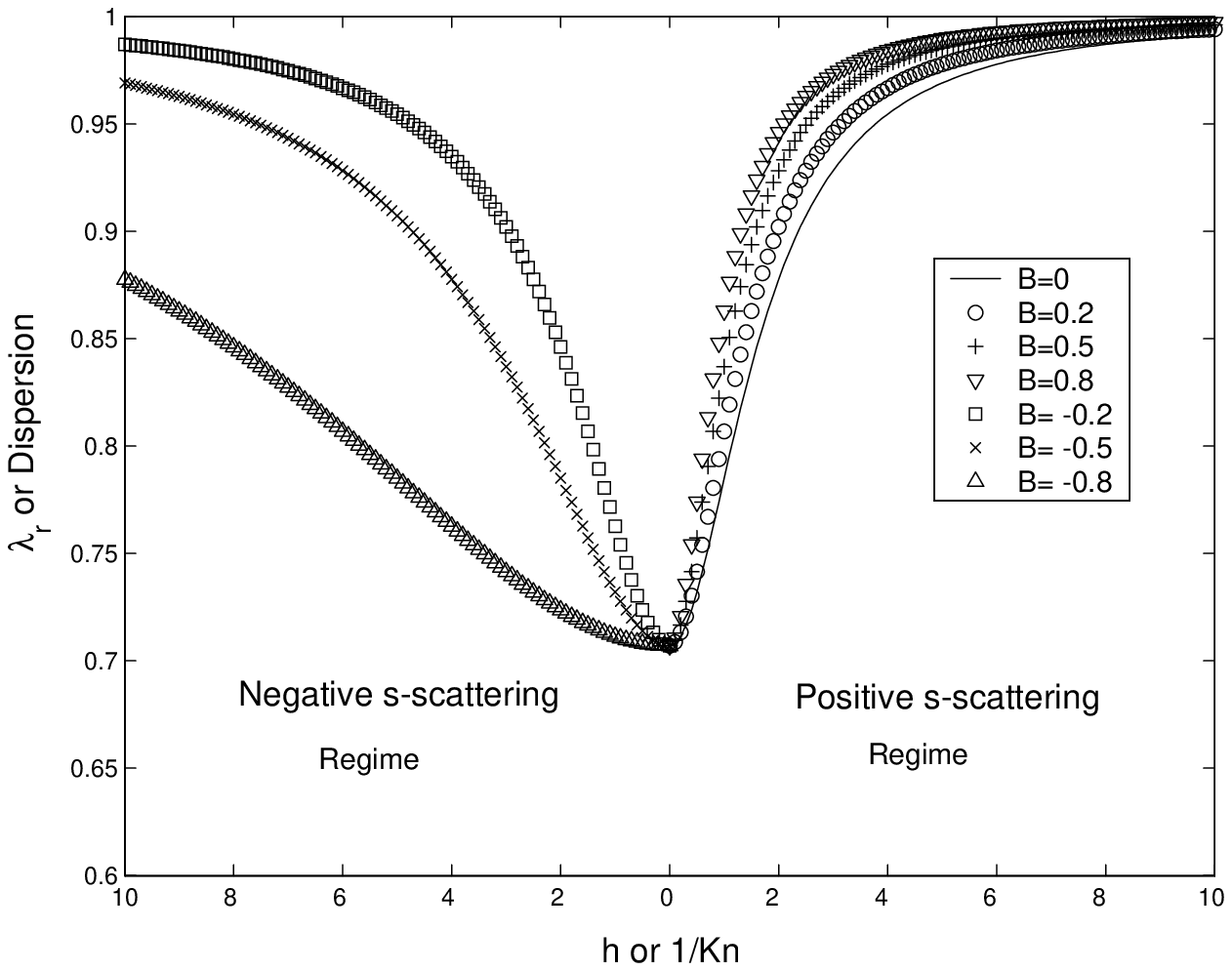,bbllx=0.2cm,bblly=11.2cm,bburx=16cm,bbury=23.2cm,rheight=10.1cm,rwidth=10.1cm,clip=}
%
\begin{figure}[h]
\hspace*{4mm} Fig. 1 \hspace*{1mm} Comparison of Bose- ($B>0$),
Boltzmann- ($B=0$), and Fermi- ($B<0$)
\newline \hspace*{4mm}  particle effects on the dispersion
($\lambda_r$). $s$-scattering means the s-wave scattering.
\newline \hspace*{4mm} The electric field is rather weak and is
neglected. The effective magnetic field ${\cal B}_{\mbox{eff}}$
\newline \hspace*{4mm} is
zero for $\nu=1/2$ in CF sense.  $B$ is the Pauli-blocking
parameter and is negative \newline \hspace*{4mm} for the case of
composite fermion gases.
\end{figure}

\newpage

\psfig{file=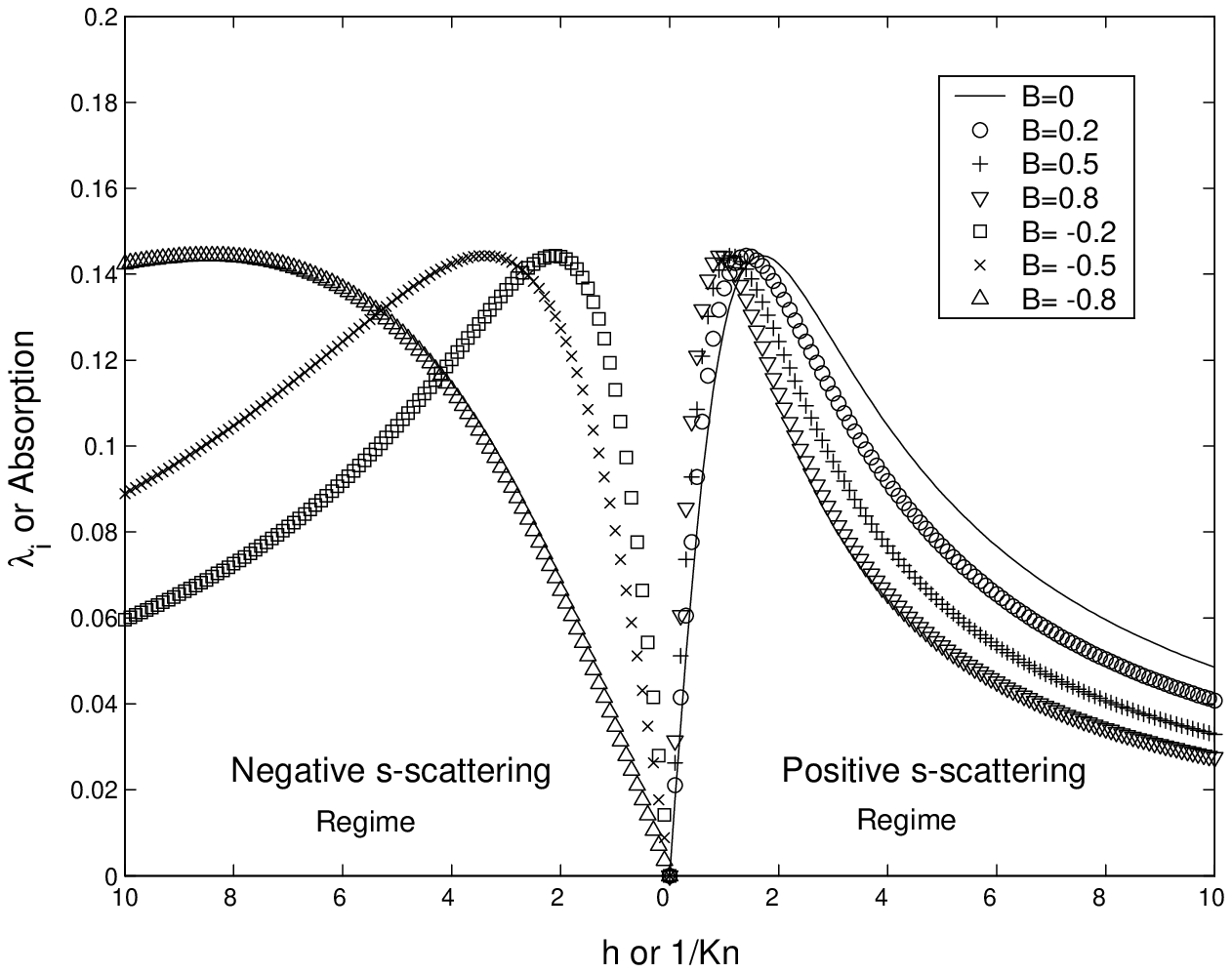,bbllx=0.2cm,bblly=11.2cm,bburx=16cm,bbury=23.2cm,rheight=10.2cm,rwidth=10.2cm,clip=}
%
\begin{figure}[h]
\hspace*{4mm} Fig. 2 \hspace*{1mm} Comparison of Bose- ($B>0$),
Boltzmann- ($B=0$), and Fermi- ($B<0$)
\newline \hspace*{4mm}  particle effects  on the absorption or attenuation ($\lambda_i$).
The rarefaction measure  $h=4 c S N_0/\omega$. \newline
\hspace*{4mm} The electric field is rather weak and is neglected.
The effective magnetic field ${\cal B}_{\mbox{eff}}$ is zero
\newline \hspace*{4mm} for $\nu=1/2$ in CF sense.
\end{figure}

\end{document}